\documentclass[prb,preprint,showpacs,amsmath,amssymb,superscriptaddress]{revtex4}

\usepackage{epsfig}
\usepackage{graphicx}

\newcommand{\beq}[1]{
\begin{equation}
\label{e#1} }

\newcommand{\eeq}{
\end{equation}
}

\begin{document}

\title{Low voltage control of ferromagnetism in a semiconductor p-n junction}

\author{M.~H.~S.~Owen}
\affiliation{Hitachi Cambridge Laboratory, Cambridge CB3 0HE, United Kingdom}
\affiliation{Microelectronics Research Centre, Cavendish Laboratory,
University of Cambridge, CB3 0HE, United Kingdom}

\author{J.~Wunderlich}
\affiliation{Hitachi Cambridge Laboratory, Cambridge CB3 0HE, United Kingdom}
\affiliation{Institute of Physics ASCR, v.v.i., Cukrovarnick\'a 10, 162 53
Praha 6, Czech Republic}

\author{V. Nov\'ak}
\affiliation{Institute of Physics ASCR, v.v.i., Cukrovarnick\'a 10, 162 53
Praha 6, Czech Republic}

\author{K. Olejn\'{\i}k}
\affiliation{Institute of Physics ASCR, v.v.i., Cukrovarnick\'a 10, 162 53
Praha 6, Czech Republic}

\author{J. Zemen}
\affiliation{Institute of Physics ASCR, v.v.i., Cukrovarnick\'a 10, 162 53
Praha 6, Czech Republic}

\author{K. V\'yborn\'y}
\affiliation{Institute of Physics ASCR, v.v.i., Cukrovarnick\'a 10, 162 53
Praha 6, Czech Republic}

\author{S. Ogawa}
\affiliation{Hitachi Cambridge Laboratory, Cambridge CB3 0HE, United Kingdom}

\author{A.~C.~Irvine}
\affiliation{Microelectronics Research Centre, Cavendish Laboratory,
University of Cambridge, CB3 0HE, United Kingdom}

\author{A. J. Ferguson}
\affiliation{Microelectronics Research Centre, Cavendish Laboratory,
University of Cambridge, CB3 0HE, United Kingdom}

\author{H.~Sirringhaus}
\affiliation{Microelectronics Research Centre, Cavendish Laboratory,
University of Cambridge, CB3 0HE, United Kingdom}

\author{T.~Jungwirth}
\affiliation{Institute of Physics ASCR, v.v.i., Cukrovarnick\'a 10, 162 53
Praha 6, Czech Republic} \affiliation{School of Physics and
Astronomy, University of Nottingham, Nottingham NG7 2RD, United Kingdom}
\date{\today}
%
%

\maketitle

{\bf
The concept of low-voltage depletion and accumulation of electron charge in semiconductors, utilized in field-effect transistors (FETs), is one of the cornerstones of current information processing technologies.
Spintronics  which is based on manipulating the collective state of electron spins in a ferromagnet provides complementary technologies for reading magnetic bits or for the solid-state memories.\cite{Wolf:2001_a} The integration of these two distinct areas of microelectronics in one physical element, with a potentially major impact on the power consumption and scalability of future devices, requires to find efficient means for controlling magnetization electrically. Current induced magnetization switching phenomena \cite{Ralph:2007_a} represent a promising step towards this goal, however, they relay on relatively large current densities. The direct approach of controlling the magnetization by low-voltage charge depletion effects is seemingly unfeasible as the two worlds of semiconductors and metal ferromagnets are separated by many orders of magnitude in their typical carrier concentrations \cite{Marder:1999_a}. Here we demonstrate that this concept is viable by reporting persistent magnetization switchings induced  by short electrical pulses of a few volts in an all-semiconductor, ferromagnetic p-n junction.}

To establish the physics behind this main result of our work, we organized the paper as follows: After describing  the basic structure of the device we present direct electrical measurements of simultaneous charge depletion and  "depletion" of the magnetization, i.e. of the shifts in the ferromagnetic Curie temperature, by applying a few volts. This provides the ultimate test of the ability to manipulate the magnetic state by low-voltages. We point out that the method we introduce for accurate electrical measurement of $T_c$ in the microdevice is essential as standard magnetometry is not feasible at these small sample dimension. We then proceed by discussing the nature of the magnetotransport responses in our system. They are remarkable on their own for their large magnitude and voltage-dependence and provide the means  for detecting the magnetic moment reorientations induced by low-voltage pulses, again by electrical measurements. The main body of the paper is devoted to presenting and discussing the phenomenology of the electrically controlled magnetization switchings measured as a function of the magnitude and angle of the external magnetic field. The microscopic physical interpretation of our experiments,  based on semiconductor theory modelling, is discussed in the concluding paragraphs.

The schematic cross-section of the III-V heterostructure used in our study is shown in Fig.1(a). It is a semiconductor p-n junction FET specially designed to accommodate ferromagnetism in the p-type region and its efficient depletion by low voltages. From the top, the structure comprises a 5~nm thick approximately 2.5\% Mn-doped GaAs capped by 2~nm of undoped GaAs to prevent oxidation of the underlying transition metal doped semiconductor film. These two top layers were grown by low-temperature molecular-beam-epitaxy (MBE) to avoid Mn precipitation. The 2.5\% doping was chosen to pass the insulator-to-metal transition threshold which for the moderately deep Mn$_{\rm Ga}$ acceptor is between 1-2\% and to achieve robust ferromagnetic state with Curie temperature $T_c\approx 30$~K, while still minimizing the number of unintentional interstitial-Mn impurities.\cite{Matsukura:2002_a,Jungwirth:2006_a,Jungwirth:2007_a} (The interstitial Mn is highly mobile at the growth temperature and its diffusion into the p-n junction would result in detrimental leakage currents.) The Curie temperature measured by SQUID in an unpatterned piece of the wafer is comparable to maximum $T_c$'s achieved at the same Mn-doping in thicker films, indicating a very good quality of our ultra-thin ferromagnetic semiconductor epilayer.

The n-type gate electrode is formed by a highly Si-doped ($2\times 10^{19}$~cm$^{-3}$) GaAs grown by high-temperature MBE. The large electron doping is required in order to achieve appreciable and voltage dependent depletion of the ferromagnetic p-region with hole doping $\sim 10^{20}$~cm$^{-3}$. The built-in electrostatic barrier due to the depletion effect at the p-n junction is further supported by inserting an undoped AlGaAs spacer with a large conduction band off-set to the n-type GaAs; for the same reason a large valence band offset AlAs spacer is placed next to the p-type (Ga,Mn)As.

Self-consistent numerical simulations,\cite{silvaco} shown in Fig.~1(b), confirm that sizable depletions are achievable by gating our heterostructure with  less than 4 Volts. Measurements discussed below were done at voltages between $-1$~V (forward bias) to $+3$~V (reverse bias) for which the leakage currents between the n-GaAs gate and p-(Ga,Mn)As channel were more than two orders of magnitude smaller than the channel currents. The (Ga,Mn)As channel was lithographically patterned in a low-resistance Corbino disk geometry with the inner contact radius of 500~$\mu$m and the outer radius of 600~$\mu$m.
\begin{figure}[h]
\epsfig{width=0.35\columnwidth,angle=-90,file=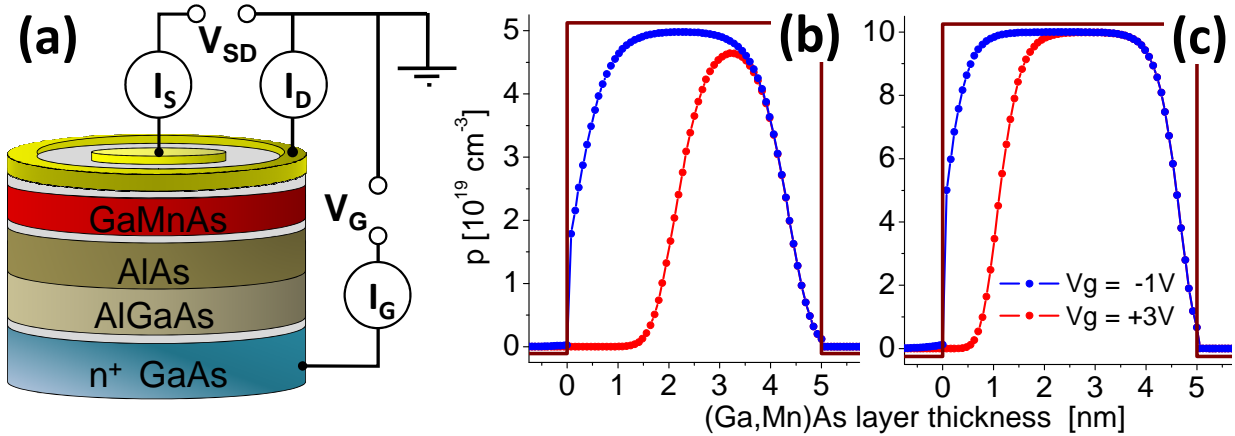}
\vspace*{0.5cm}
\caption{(a) Schematic of the ferromagnetic p-n junction FET structure and the Corbino disk geometry of the source and drain contacts. (b) and (c) Numerical simulations of the hole density profile at -1~V (accumulation) and +3~V (depletion), considering a $2\times 10^{19}$~cm$^{-3}$ electron doping in the n-GaAs and $5\times 10^{19}$~cm$^{-3}$ and $10^{20}$~cm$^{-3}$ hole doping in the p-(Ga,Mn)As.
}
\label{f1}
\end{figure}

In Fig.~2(a) we plot the measured channel resistances vs. gate-voltage  at temperature 4-40~K. Both at low temperatures and near $T_c$ we observe a marked increase of the channel resistance at positive voltages. It is consistent with the depletion of the (Ga,Mn)As channel as predicted by the simulations in  Fig.~1(b) and with the vicinity of the metal-insulator transition which causes the superlinear increase of $R$ with $V_g$. At 4~K, the increase of $R$ between -1 and +3~V is by more than 100\%.

In Fig.~2(b) we show the voltage-dependence of the Curie temperature in the ferromagnetic p-n junction. Our measurement technique is distinct from previous studies which relied on approximate extrapolation schemes based on Arrot plot measurements at finite magnetic fields.\cite{Ohno:2000_a,Chiba:2006_b,Stolichnov:2008_a} Recent observation and interpretation by the authors \cite{Novak:2008_a} of the peak in the zero-field temperature derivative of the resistance at the Curie point  in good quality (Ga,Mn)As materials has provided the tool for direct transport measurements of $T_c$ in microdevices without relying on any extrapolation schemes. In Fig.~2(b) we plot differentiated resistivity curves obtained in our device at -1 and +3~V. The data show a clear shift of the Curie temperature, i.e., the magnetization can be turned on and off in parallel with accumulating and depleting holes in the ferromagnetic semiconductor channel by biasing the p-n junctions with a few Volts.

Curie temperature variations provide the key physical demonstration of the low-voltage control of magnetization. Nevertheless, for most spintronic functionalities it is not required to destroy the ordered state of spins but only to change their collective orientation. We therefore focus on effects related to reorientations of the unit vector of the macroscopic moment. To avoid thermal fluctuations of the magnetization, all measurements are done far from the Curie point at 4~K.

In Fig.~2(c) we show magnetoresistance traces recorded during in-plane and perpendicular-to-plane sweeps of an external magnetic field, at gate voltages of -1 and +3~V. Apart from the negative isotropic magnetoresistance (IMR), the data indicate a remarkably large anisotropic magnetoresistance (AMR) effect which at saturation reaches $\sim 30$\%. (Note that AMR sensors fabricated in transition metal ferromagnets with AMR ratios of a few per cent \cite{McGuire:1975_a} marked the dawn of spintronics in the early 1990's.) The resistance is larger for the  perpendicular-to-plane magnetization orientation and the size of the effect is enhanced by depletion. The electrical response of our system to magnetization rotations is both large and tuneable by low gate voltages.

The magnetoresistance traces in Fig.~2(c) indicate that the film has a magnetic anisotropy favoring in-plane magnetization, which is overcome by an external field of approximately 150~mT. At weaker magnetic fields, magnetization switching effects are confined to the plane of the ferromagnetic film. The qualitative nature of the in-plane magnetic anisotropy landscape which determines the switching processes can be scanned in our Corbino microdevice by recording the AMR at a rotating in-plane saturation field. Unlike in the out-of-plane rotation AMR, contributions depending on the relative angle between the in-plane magnetization and current average out over the radial current lines. The in-plane AMR then depends purely on the angle between magnetization and crystallographic axes.\cite{Rushforth:2007_a} It reflects therefore the same underlying symmetry breaking crystal fields as the magnetic anisotropy. The measurements, shown in Fig.~2(d), unveil a cubic anisotropy along the [110]/[1$\bar{1}$0] crystal axes and an additional uniaxial term breaking the symmetry between the [110] and [1$\bar{1}$0] directions. Although the specific responses to these symmetries can be very different for the AMR and for the magnetic anisotropy, the presence of the cubic and uniaxial AMR terms and their sensitivity to the gate voltage observed in Fig.~2(d) suggest that the in-plane magnetization orientation itself can be switched at weak magnetic fields by the low voltage charge accumulation or depletion.
\begin{figure}[h]
\epsfig{width=.7\columnwidth,angle=-90,file=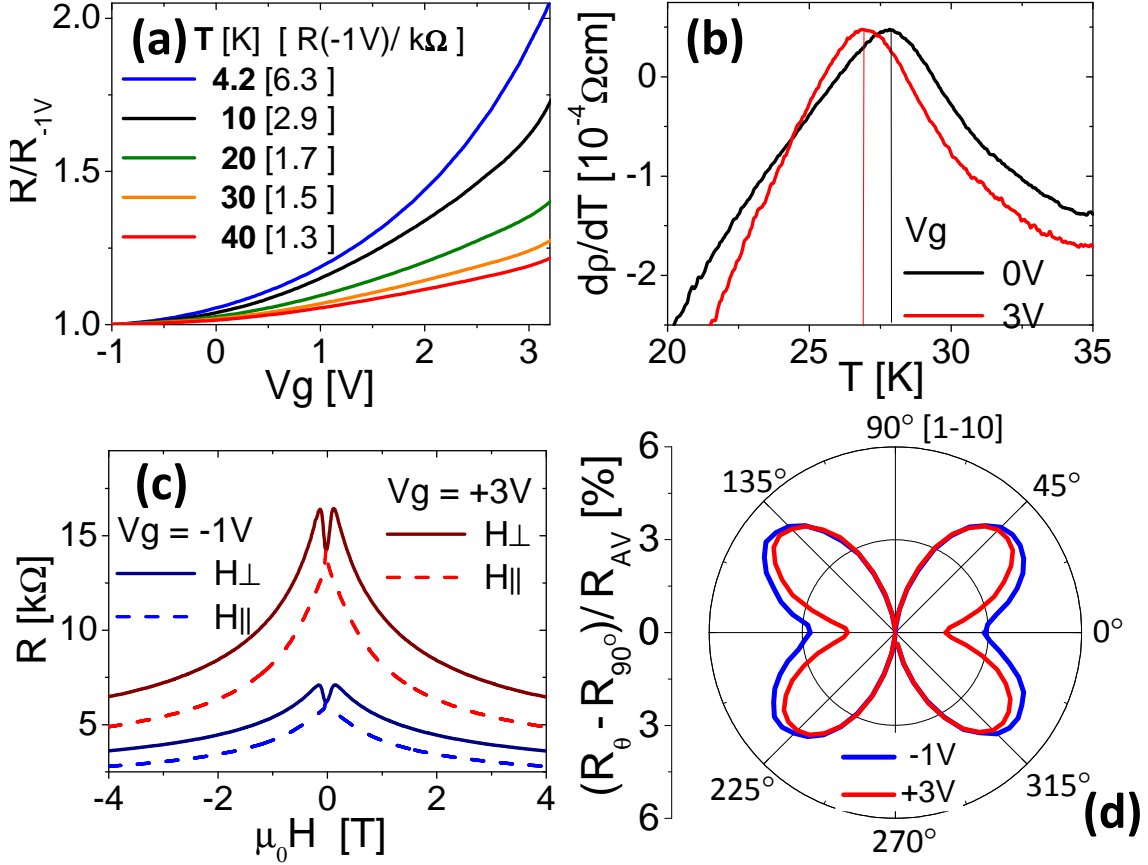}
\vspace*{0.5cm}
\caption{(a) Gate voltage dependence of the p-(Ga,Mn)As channel resistance at temperatures 4-40~K. (b) Temperature derivative of the measured channel resistivity at 0 and +3~V. The maximum corresponds to $T_c$. (c) In-plane ($H_{\parallel}$) and perpendicular-to-plane ($H_{\perp}$) magnetic field sweep measurements of the channel resistance at -1 and +3~V. The difference between the $H_{\parallel}$ and $H_{\perp}$ sweeps above the reorientation field of $\approx 150$~mT corresponds to the out-of-plane AMR. (d) In-plane AMR measured at saturation in a rotating in-plane field at -1 and +3~V.
}
\label{f2}
\end{figure}

A variable width of hysteretic magnetization loops measured at different constant gate voltages,  shown in Fig.3(a), is the prerequisite for observing electrically assisted magnetization switchings. Note that electrical measurements of magnetization reorientations utilized in Figs.~3 and 4 are facilitated in our system by the IMR which responds to abrupt changes of the total magnetic induction upon a 180$^{\circ}$ reversal, and by a combined effect of the IMR and of the AMR for intermediate switchings by less than 180$^{\circ}$. The amplitude of the AMR and the IMR contributions are similar in our experiments. The switchings by short low-voltage pulses are demonstrated in Fig.~3(b) and analyzed in detail in Fig.~4. The experiments were performed at constant field-sweep rate of 0.1~mT per second starting from negative saturation field of 1~T. The gate voltage was set to a base value of -1~V and then after each measurement step spanning 1 second we applied a 10~ms voltage pulse of a fixed magnitude and then returned to the base voltage. The technique allows us to demonstrate magnetic response to short electric pulses and the persistence of induced reorientations of the magnetization vector. It also removes potentially obscuring variations among the resistance traces in regions away from magnetization switchings which are caused by different slopes of the  negative IMR at different gate voltages.

In Fig.~3(b) we compare measurement with no pulses (constant -1~V gate voltage) and data acquired at 0 to +4~V peak voltages. The  field was swept along the [1$\bar{1}$0] crystal direction ($\theta = 90^{\circ}$, where $\theta$ is the in-plane field angle measured from the [110] direction). As argued in detail in Fig.~4 and confirmed by SQUID magnetization measurements on an unpatterned part of the wafer, [1$\bar{1}$0] is the main magnetic easy axis. The negative IMR then allows  us to observe the drop in $R$ corresponding to a 180$^{\circ}$ reversal from an antiparallel to a parallel configuration of field and magnetization and a corresponding increase of the magnetic induction. As the applied peak voltage increases, the magnetization reversals consistently shift to lower magnetic fields and the magnetization remains switched when the peak voltage pulse is turned off.

\begin{figure}[h]
\vspace*{1cm}
\hspace*{-1cm}\epsfig{width=.35\columnwidth,angle=-90,file=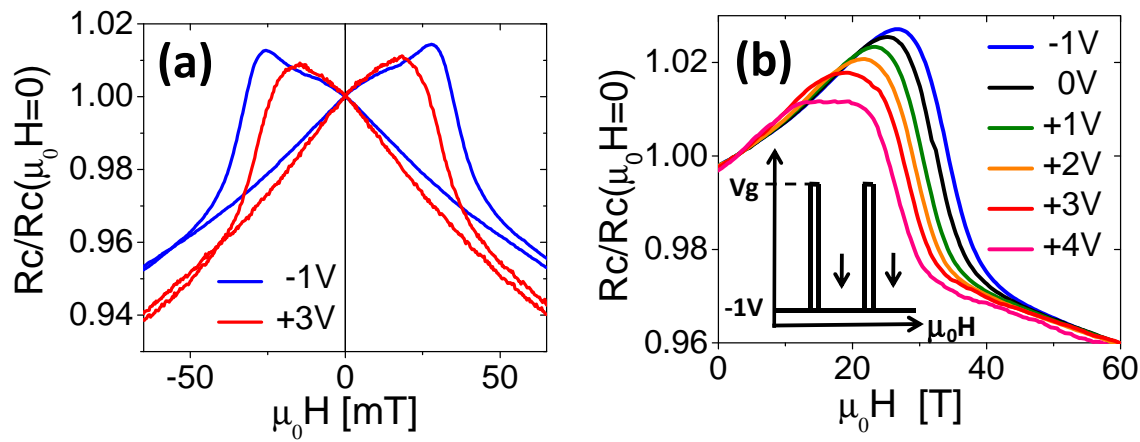}
\vspace*{0.5cm}
\caption{(a) Hysteretic field sweep measurements at field angle $\theta=90^{\circ}$ for constant gate voltages of -1 and +3~V. (b) Up-sweeps of the  $\theta=90^{\circ}$ in-plane field at constant -1V gate voltage, and for measurements with the gate voltage set to a base value of -1~V and with short additional voltage pulses corresponding to a total peak voltage of 0, +1 , +2, +3, and +4~V, respectively.
}
\label{f3}
\end{figure}

To discuss the detail phenomenology of these persistent low-voltage induced magnetization switchings we present in Figs.~4(a) and (b) field-sweep measurements at fixed field angles spanning the whole interval from 0 to 180$^{\circ}$ in 5$^{\circ}$ steps. In panels (a) and (b) we show color-maps of the resistance as a function of the field magnitude and angle for -1~V constant voltage and for the +3~V peak-voltage measurements, respectively. The main effect observed in these  plots is the overall suppression of the magnitude of the switching fields by depletion. Additionally, the relative suppression is stronger at $\theta=0$ than at  90$^{\circ}$, as highlighted in Fig.~4(c). This indicates that both the magnitude and ratio between the uniaxial and cubic anisotropy fields is modified by the gate voltage. To quantify the depletion induced modification of the magnetic anisotropy we extracted the anisotropy constants from fitting the measured $\theta=0$ and 90$^{\circ}$ switching fields to a single domain anisotropy energy model, $E(\theta,\phi)=K_u\sin^2\theta-K_c\sin^22\theta/4-MH\cos(\phi-\theta)$, where $H$ and $M$ are the magnitudes of the external field and magnetization, respectively. The uniaxial constant $K_u$ is relatively weak compared to the cubic constant $K_c$, as shown in Fig.~4(d). They both have a negative sign corresponding to the magnetic easy directions along the [1$\bar{1}$0] and [110] axes and the most easy direction along  [1$\bar{1}$0].  As shown also apparent from Fig.~4(d), the dominant effect of depletion is in reducing the magnitude of $K_c$. Fig.~4(e) shows how the corresponding anisotropy energy profiles at $H=0$ evolve with depletion.
\begin{figure}[h]

\vspace*{1cm}
\epsfig{width=0.67\columnwidth,angle=90,file=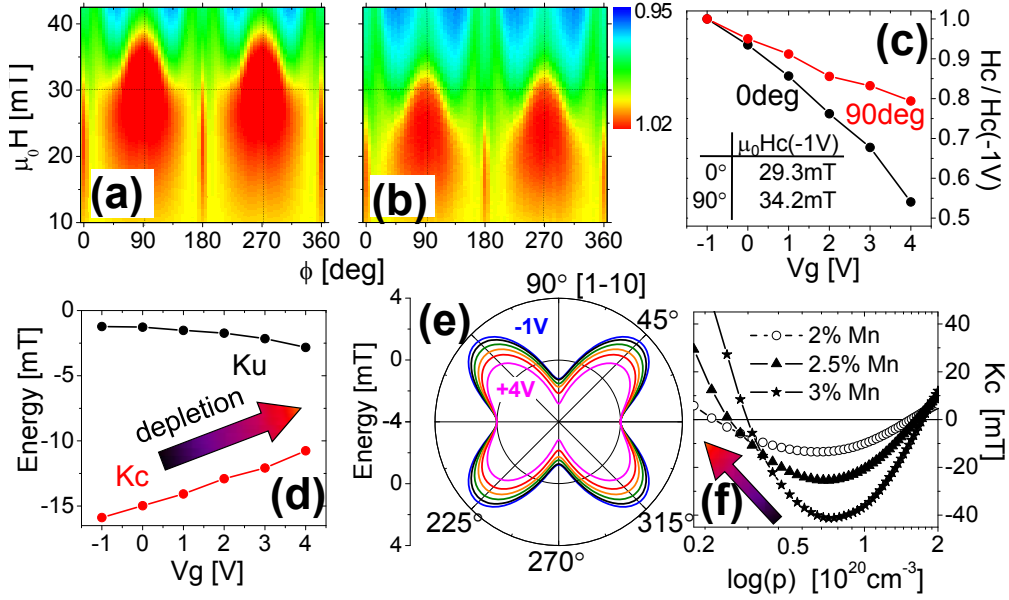}
\vspace*{0.5cm}
\caption{(a) and (b) Color maps of channel resistance as a function of the in-plane field angle and magnitude (normalized to $H=0$ resistance) for -1V constant voltage and +3V peak voltage measurements, respectively. (c) Switching fields at field angles $\theta=0$ than at  90$^{\circ}$ as a function of the gate voltage. (d) Uniaxial and cubic anisotropy constants and (e) corresponding anisotropy energy profiles derived from the measured $\theta=0$ than at  90$^{\circ}$ switching fields. (f) Microscopic calculations of the cubic anisotropy constant. Arrows in (d) and (f) highlight the common trend with depletion.
}
\label{f4}
\end{figure}

We now discuss the key experimental observations by employing the ${\bf k}\cdot{\bf p}$ semiconductor theory approach combined with the mean-field kinetic-exchange model of hole mediated ferromagnetism in (Ga,Mn)As.\cite{Matsukura:2002_a,Jungwirth:2006_a}
Calculations for 2.5\% local moment doping and  hole density $p\sim 1\times 10^{20}$~cm$^{-3}$, for which the simulations in Fig.1(b) predict hole depletions consistent with the measured variations of the channel resistance at temperatures near $T_c$, yield $T_c\sim 20~K$ and $dT_c/dp\approx 1\times 10^{-19}$~Kcm$^{3}$. Both the absolute value of the Curie temperature and the few Kelvin suppression of $T_c$ at a $\sim 20$\% hole depletion predicted by the theory are consistent with our p-n junction simulations and the measured gate-dependent $T_c$ values.

The semiconductor theory modelling which includes strong spin-orbit coupling effects in the host semiconductor valence band captures also the sensitivity of magnetocrystalline anisotropies in (Ga,Mn)As to hole density variations. The cubic anisotropy is included by accounting in the  ${\bf k}\cdot{\bf p}$ model for the zincblende crystal structure of GaAs. The additional weak uniaxial anisotropy is often present in (Ga,Mn)As epilayers but its microscopic origin is not known and we will therefore focus only on the stronger cubic anisotropy term. As shown in Fig.~4(f), the microscopically calculated $K_c$ constant changes sign at hole density of approximately $1.5\times 10^{20}$~cm$^{-3}$. Below this density it favors the [110]/[1$\bar{1}$0] magnetization directions, consistent with the experimental data. The typical magnitudes of $K_c$ of $\sim 10$~mT are also consistent with experiment and considering the large gate action seen at low temperatures we can also associate, semiquantitatively, the decreasing magnitude of the experimental $K_c$ at depletion with the behavior of the theoretical $K_c$ at low hole densities.

To conclude we have reported low-voltage control of magnetic properties of a p-n junction FET via depletion effect in the ferromagnetic semiconductor channel. We have shown variable $T_c$ and AMR, and demonstrated magnetization switchings induced by short electric field pulses of a few volts. Our concept of the spintronic transistor  is distinct from previously demonstrated high-voltage metal-oxide-semiconductor ferromagnetic FETs\cite{Ohno:2000_a,Chiba:2006_b,Stolichnov:2008_a} or electro-machanically gated ferromagnets by piezo-stressors.\cite{Kim:2003_a,Lee:2003_b,Botters:2006_a,Boukari:2007_a,Rushforth:2008_a,Overby:2008_a,Goennenwein:2008_a}
It is realized in an all-semiconductor epitaxial structure and offers a principally much faster operation. In basic physics research, we expect broad utility of our results in studies of carrier-mediated ferromagnetism and in interdisciplinary fields combining ferromagnetism and spin-orbit coupling effects with localization and quantum-coherent transport phenomena\cite{Neumaier:2007_a} controlled by carrier depletion.

We acknowledge helpful discussions with R.~P.~Campion, M.~Cukr, B.~L.~Gallagher, M.~Mary\v{s}ko, J. Sinova, and J.~Zemek, and from EU Grant IST-015728, from Czech Republic Grants FON/06/E001, FON/06/E002, AV0Z1010052, KAN400100652, and LC510, and from U.S. Grant SWAN-NRI.


\end{document}